\documentclass[sigconf]{acmart}

\usepackage[english]{babel}
\usepackage{svg}
\usepackage{pgfplots}
\usepgfplotslibrary{dateplot}
\usepgfplotslibrary{colorbrewer}
\pgfplotsset{cycle list/Dark2}
\usepackage{tikz}
\usetikzlibrary{calc,positioning,arrows,shapes,shapes.misc,pgfplots.dateplot}
\usepackage{pgfplotstable}
\usepackage{pgfcalendar}
\usepackage{csquotes}
\usepackage[htt]{hyphenat}

\begin{document}

\acmConference[]{}{}{}
\settopmatter{printacmref=false} 
\renewcommand\footnotetextcopyrightpermission[1]{} 
\setcopyright{none}

\title{Observing the Evolution of QUIC Implementations}

\author{Maxime Piraux, Quentin De Coninck, Olivier Bonaventure}
\affiliation{%
  \institution{UCLouvain}
}
\email{{maxime.piraux, quentin.deconinck, olivier.bonaventure}@uclouvain.be}

\setlength{\belowcaptionskip}{-11pt}

\begin{abstract}
The QUIC protocol combines features that were initially found inside the TCP, TLS and HTTP/2 protocols. The IETF is currently finalising a complete specification of this protocol. More than a dozen of independent implementations have been developed in parallel with these standardisation activities.

We propose and implement a QUIC test suite that interacts with public QUIC servers to verify their conformance with key features of the IETF specification. Our measurements, gathered over a semester, provide a unique viewpoint on the evolution of the QUIC protocol and of its implementations. They highlight the introduction of new features and some regressions among the different implementations.
\end{abstract}

\maketitle

\section{Introduction}

Internet transport protocols usually evolve slowly. Any significant evolution to TCP, the dominant transport protocol, takes years of efforts to be widely deployed. There are several factors that explain this slow evolution \cite{papastergiou2017ossifying}. On one hand, TCP is usually implemented inside the operating system kernel and upgrading kernels remains costly and slow. On the other hand, there are a growing number of middleboxes on the Internet that block new TCP extensions \cite{honda2011still}. QUIC, initially proposed by Google \cite{langley2017quic} addresses this ossification in several ways. First, QUIC runs above UDP, which implies that it can be distributed as a userspace library that can be easily upgraded. Second, QUIC encrypts most of the control information and payload in order to prevent middlebox interferences.

The results obtained by Google with QUIC\,\cite{langley2017quic} combined with its security features have convinced the IETF to standardise a new protocol starting from Google's initial design. The QUIC IETF working group was chartered in late 2016. It is currently finalising the specifications for the first standards-track version of QUIC. During this period, the IETF working group published fourteen versions of the main protocol draft and more than a dozen of QUIC implementations are publicly available.

\begin{figure}
	\centering
	\includegraphics[width=\columnwidth]{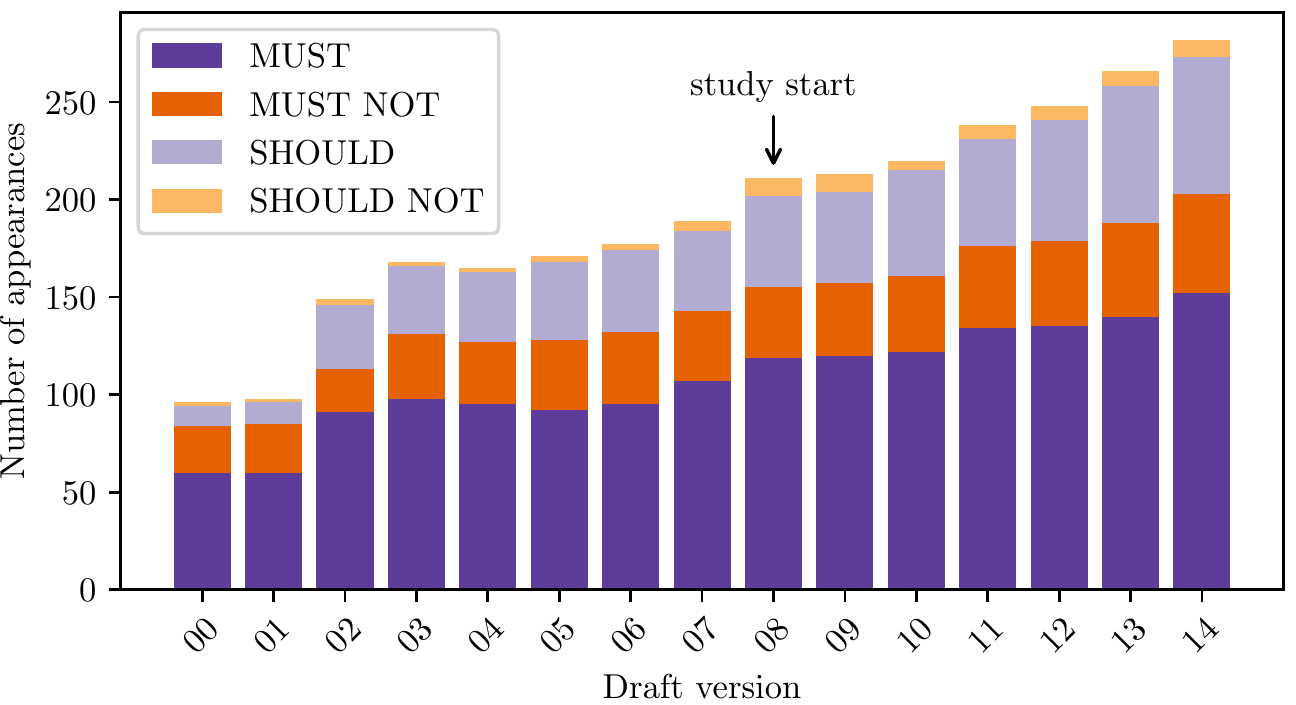}
	\caption{Keywords in \texttt{draft-ietf-quic-transport}.}
	\label{fig:requirements}
\end{figure}

The efforts of the QUIC designers and implementers is probably unique in the history of protocol design and implementation. Although many IETF protocols have been designed, very few have attracted so many implementers while the protocol was still being developed. As an illustration of the complexity of the QUIC protocol, Figure~\ref{fig:requirements} plots the number of RFC2119 keywords (i.e.  \enquote{MUST}, \enquote{MUST NOT}, \enquote{SHOULD}, \enquote{SHOULD NOT}) in the different versions of \texttt{draft-ietf-quic-transport}. We can observe that the number of these indicators has more than doubled since the first version of the specification.

Implementing network protocols is not a trivial task and several authors have proposed techniques to test and validate protocol implementations. Some of these techniques rely on formal methods to automatically derive the test suite from the implementation\,\cite{grabowski2003introduction,holzmann1991design}. However, it is difficult to apply them to Internet protocols since their specifications are informal. Researchers have proposed different solutions to test and validate protocol implementations. Some have proposed techniques to validate complete TCP implementations\,\cite{musuvathi2004model,bishop2005rigorous}. Paxson et al.\ proposed specific tools to validate the conformance of TCP implementations\,\cite{paxson1997automated,RFC2525}. With TBIT, Padhye and Floyd developed techniques to infer the characteristics of TCP implementations by interacting with them with specially crafted packets\,\cite{pahdye2001inferring}. These tools have played an important role in improving TCP implementations.

The dozen QUIC implementations\,\cite{quic-implem} that are actively being developed will likely face similar problems as TCP implementations during the last decades\,\cite{RFC2525}. The interoperability tests that are regularly organised by the QUIC IETF working group have helped to identify some ambiguities in the specifications and solve interoperability problems. In this paper, we contribute to this implementation effort with a publicly available and detailed test suite for QUIC. To our knowledge, this is the first public test suite for this new protocol.

We first describe the architecture of our test suite in Section~\ref{sec:methodology}. Section~\ref{sec:results} analyses results collected during a 6-month period using our test suite on the public servers that already implement QUIC. Section~\ref{sec:discussion} concludes this paper by reviewing the future prospects for our work and assessing how it can be freely improved, reused and extended.

\section{The QUIC Test Suite}\label{sec:methodology}

In this section we first describe both the approach and the architecture of our test suite. We then depict the test scenarios that constitute it.

\subsection{Testing approach}

Network protocol testing approaches can be categorised according to two dimensions, black-box versus white-box testing and passive versus active testing. The first dimension defines the perspective chosen to evaluate an implementation, i.e.\ an external or internal perspective. Because we want to test all QUIC implementations without relying on source code availability, we chose the black-box approach. Our test suite only observes their external behaviours, i.e.\ the packets sent and receiver, to evaluate them.

The second dimension defines how the tool behave with respect to the implementation under test (IUT).
The first approach is passive testing.
It has been used in earlier works\,\cite{paxson1997automated, rewaskar2006passive, jaiswal2004inferring}, but is of limited use with QUIC given that most of the packets are encrypted.

The second approach is active testing, in which the tool used for experiments actively exchanges messages with the IUT. It requires the IUT to be available when using the tool. Several studies have applied this approach to TCP. TBIT \cite{pahdye2001inferring} was one of the pioneering work in this domain. It has been extended in later works\,\cite{ekiz2011misbehaviors, yang2014tcp}. Conducting active tests with TCP is becoming more difficult today given the deployment of various types of middleboxes that may interfere with active tests \cite{honda2011still,medina2004measuring,hesmans2013tcp}. By encrypting most of the control information and payload, QUIC exposes a smaller surface subject to ossification.

The objective of our test suite is to check the conformance of QUIC server-side implementations by only exchanging packets with them. Given that the IETF specifications are still being developed, we only cover a subset of them.
We want the tool to be autonomous in two manners. It must be able to create QUIC packets on its own to perform the tests. It must also be able to appropriately present bug reports for an implementer to locate which mechanisms caused their occurrence.

\subsection{Architecture}

\begin{figure}
    \centering
    \begin{tikzpicture}[
        node distance=1.5cm,
        >=stealth',
        auto,
        font=\small,
        <->
    ]
    \tikzstyle{comp}=[draw, rectangle, minimum height=0.7cm] 
    
    \node[comp] (imp) {Implementations};
    \node[above of = imp,comp] (toolbox) {\textbf{QUIC toolbox}};
    \node[above of = toolbox,comp] (testsuite) {\textbf{Test scenarii}};
    \node[right = of testsuite,comp] (traces) {\textbf{Traces}};
    \node[right = of traces,comp] (dissector) {\textbf{Dissector}};
    \node[below of = dissector,comp] (webapp) {\textbf{Web site}};
    \node[below of = webapp,comp] (implementers) {Implementers};
    
    \draw[<-] (imp) -- node[midway, right, align = left]{Exchange QUIC packets \\ over the Internet} (toolbox);
    \draw[<-] (toolbox) -- node[midway, right, align = left]{Call} (testsuite);
    \draw[->] (testsuite) -- node[midway, above]{Produce} (traces);
    \draw[<-] (traces) -- node[midway, above]{Reads} (dissector);
    \draw[<-] (dissector) -- node[midway, left, align = right]{Calls} (webapp);
    \draw[<-] (webapp) -- node[midway, left]{Consult } (implementers);
    
    \end{tikzpicture}
    \caption{Tools forming the test suite.}
    \label{fig:architecture}
\end{figure}
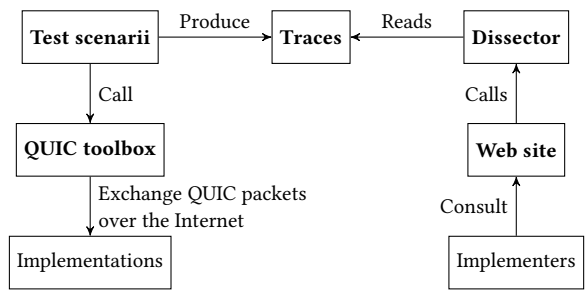

Figure~\ref{fig:architecture} illustrates the architecture of our test suite.
Akin to TBIT\,\cite{pahdye2001inferring}, our test suite is constituted of several self-contained scenarios. Each scenario is self-contained in the sense that it establishes a new QUIC connection to perform the test. A separate connection increases the isolation between two scenarios, in an attempt to accurately test a specific mechanism.  Each test addresses a particular feature of the QUIC protocol. Combining several scenarios within a single connection is left as future work.

We implement the scenarios on top of a high-level API, i.e.\ the QUIC toolbox, that allows easily manipulating QUIC packets. Pieces of QUIC client behaviour are implemented as asynchronous message passing objects, called agents. We implemented 9 different agents responsible for, e.g.\ issuing \texttt{ACK} frames in response of received packets, retransmitting lost data, interacting with TLS, decrypting and parsing QUIC packets, bundling frames into packets and performing 1-RTT handshake.

This increases modularity by defining the behaviour of a test without having to reimplement  mechanisms shared by several tests. For instance, the \texttt{address\_validation} scenario, which tests whether a server validates the client address before sending significant amount of data to it, does not send acknowledgements but retransmits lost data and derive session keys from the handshake to decrypt all received packets. Therefore, it disables the agent responsible for acknowledgements, but enables the agents interacting with TLS and sending retransmissions.

The QUIC toolbox depends on \textit{picotls}\,\cite{picotls} for using TLS 1.3 through a Go binding we wrote\,\cite{pigotls}. The interaction between TLS and the toolbox is isolated in a separate 115-line long agent. One can thus replace the TLS stack used by implementing a new agent providing equivalent functionalities.

We synchronise the different agents using specific events inside a connection, e.g. a packet has been received or sent out, a new encryption scheme is available. This paradigm allows attaching new behaviours upon reception of these events without requiring extra coordination with other agents or having to define a common event loop for each connection. Agents can also emit events as their connection progresses.

Each test outputs its result in a JSON trace. Each trace contains an error code that summarises its outcome and a clear-text trace of the packets exchanged during its execution. The error code is not purely binary, i.e.\ passed or failed. It can be used to discern various cases of failures to help the implementer to locate which part of the tested mechanism was deemed as erroneously implemented. For instance, the \texttt{zero\_rtt} test can report whether a valid resumption token was sent by the server, whether it allowed the test suite to effectively succeed a 0-RTT handshake and whether the test suite could perform an 0-RTT HTTP request.

A trace can also contain scenario-specific data, such as the list of the protocol versions that were announced during the \texttt{version\_negotiation} test and the transport parameters received during the \texttt{transport\_parameters} test. Using this feature, we instrumented several test scenarios to collect several metrics. We present three of them in the Section~\ref{sec:results}.

To track the evolution of QUIC implementations, we ran an instance of the test suite every day. The different scenarios are run in a randomised order to prevent a particular sequence of tests from repeatedly impacting the data collection. A publicly accessible web application allows visualising the test results\,\cite{quic-tracker-web-site}. It eases the communication of bug reports to implementers and hopefully encourages them to consult test results. The presentation of the results emphasises on the cause of the problem so that implementers can efficiently diagnose which mechanism was erroneously implemented. The website also provides a detailed description of each test.

Our web application embeds a packet dissector we implemented. We chose to develop our own because existing packet dissectors, such as those in Wireshark\,\cite{wireshark-quic}, do not consistently support QUIC. Being able to dissect the packets exchanged by the test suite greatly improves its ability to efficiently present bug reports. The dissector operates based on a specification of the protocol written in YAML and a cleartext trace of the packets exchanged. We maintain separate specifications for different QUIC versions in order to ensure backward compatibility.

Our QUIC toolbox consists of more than 3600 lines of Go code. The web application consists of 1100 lines of Python code, while the dissector is 300-line long supplemented by 1600 lines of YAML for protocol descriptions.

\subsection{Testing the specification}

We derive test scenarios from the QUIC specification. This process cannot be automated, because the specification is written in English in an informative style.
We analyse the sentences containing strong indicators of importance as defined in RFC2119\,\cite{RFC2119}, i.e.\ sentences containing the words \enquote{MUST} or \enquote{MUST NOT}. We then extract rules from these sentences that should not be violated throughout the test. Based on these rules we design a scenario that ensures that these rules are not violated. The tests follow the evolution of the specification and are updated accordingly.
We prioritise the design of tests that involve features chosen by the QUIC working group as part of the \textit{Implementation Drafts}\,\cite{7th-implementation-draft} to provide valuable feedback.

Our current test suite contains 18 test scenarios. In the interest of space, we do not present all of them but summarise the mechanisms they verify. Eight of them check several aspects of the QUIC handshake, such as the \textit{0-RTT}, the exchange of the \textit{Transport Parameters} and the \textit{Version Negotiation}. Six of them focus on QUIC streams, e.g. bidirectional and unidirectional support, flow control and reordering in stream transitions. Two of them test the handling of acknowledgements and the support for Explicit Congestion Notification (ECN). Finally, two tests explore connection migration and new connection ID support.

By manually analysing the 14th version of \texttt{draft-ietf-quic-transport}, we identified 29 strong requirements covered by the test suite, i.e.\ \enquote{MUST} and \enquote{MUST NOT}. We note that 41 of the 203 strong requirements are only applicable to QUIC client implementations and thus out of the scope of our tool. These 18 test scenarios are implemented in about 1200 lines of Go code.

\section{Test Results}\label{sec:results}

We used our QUIC test suite on the public endpoints of QUIC implementations during a 6-month period, starting from the 12th of February to the 15th of July 2018\footnote{A bug was introduced on the 1st of May, preventing data collection until the 8th of May.}. We updated the list of public endpoints when they were publicly announced on the communication channel dedicated to testing coordination\,\cite{quicdev-slack}.

We report our results into three phases.
First, we provide a high-level view showing key metrics collected by our test suite.
Then, we dig in two case studies on two specific test scenarios.
Finally, we present a snapshot of the test results to show the diversity of behaviours between studied implementations.

\subsection{Measurements}\label{sec:measurements}

In this section, we present three metrics extracted from the data collected by the test suite, i.e.\ the announced QUIC version, the handshake success and the test outcome percentage. For each metric we explain how the measurements were conducted and what conclusion can be drawn from them.

\begin{figure}
	\centering
	\includegraphics[width=\columnwidth]{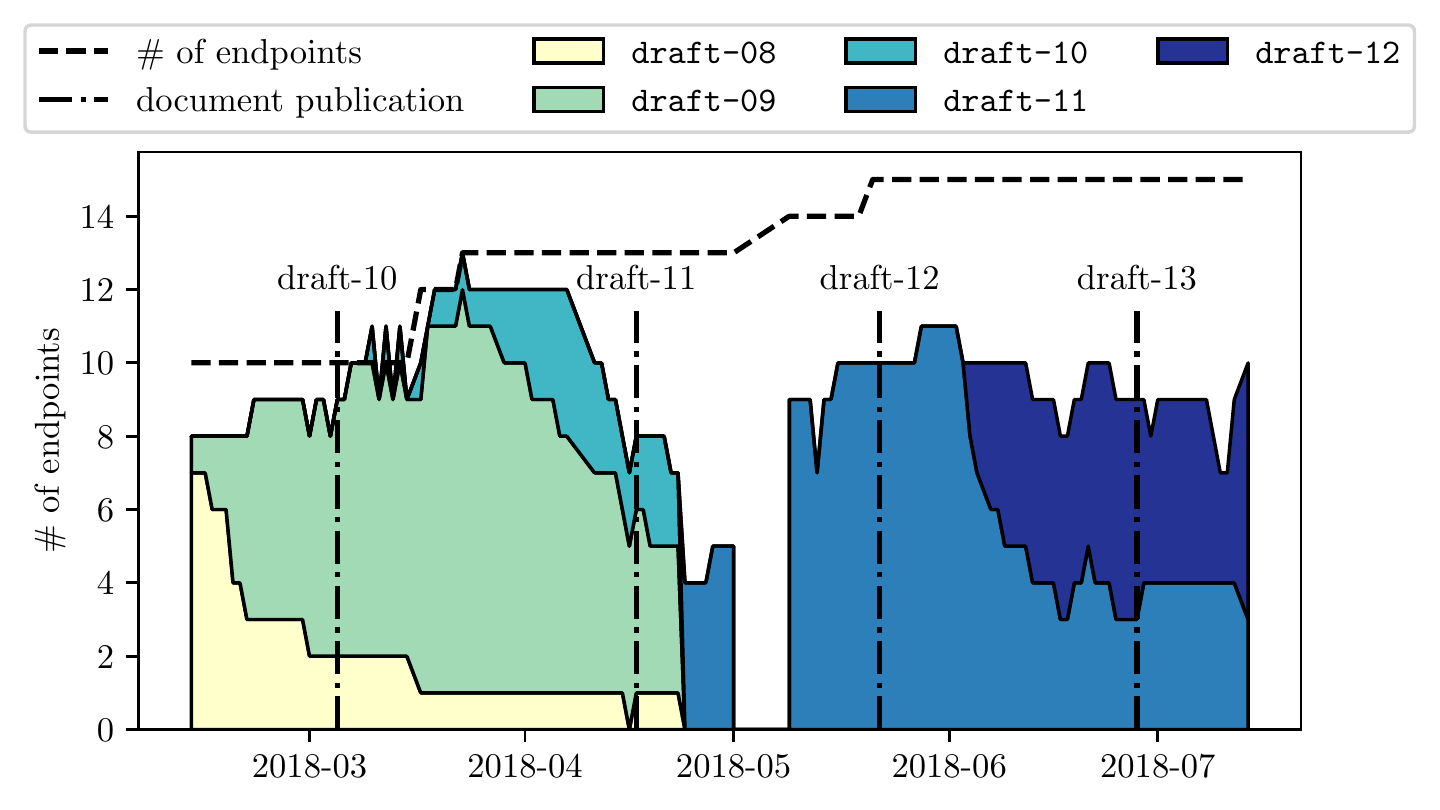}
	\caption{Number of endpoints announcing different draft versions.}
	\label{fig:draft-versions}
\end{figure}

\textbf{Deployment of QUIC.}
We first report the result collected by the \texttt{version\_negotiation} scenario. This test triggers the version negotiation process and records the versions announced by the tested implementations. In a sense, this is similar to the measurements carried out by R{\"u}th et al.\ to identify the number of servers that support Google's version of QUIC (gQUIC) over the entire IPv4 addressing space\,\cite{quicWildPAM2018}. Figure~\ref{fig:draft-versions} summarises our results over the 6-month period. We restrict the figure to the five main versions of the draft specifications\,\cite{quic-versions}. It also indicates the number of endpoints we tested over this period.

We can observe that when a new version of the specification was published, most implementations chose to stop supporting the previous version in favour of the new one without maintaining backward compatibility. This is reflected in the figure by a simultaneous increase and decrease between two successive versions. This lack of backward compatibility is normal for prototype implementations.
It contrasts with the findings of R{\"u}th et al.\ about public gQUIC servers that gradually update the set of versions they support.

QUIC versions can introduce lot of changes, including modifications to the QUIC invariants about the public header format.
\texttt{draft-11} is an example. It introduced a version negotiation process that is not backward-compatible. We updated the test suite to support \texttt{draft-11} shortly after its publication. From this point, we were unable to observe prior versions.

We can conclude from Figure~\ref{fig:draft-versions} that the implementers of QUIC often need between fifteen days to a month to integrate the changes published in a new version of the specification to their implementations. As a result, tracking the behaviour of these QUIC implementations requires to regularly update the test suite.

\begin{figure}
    \centering
    \begin{tikzpicture}[node distance=0.5\columnwidth,auto,>=stealth']
    \node[] (server) {Server};
    \node[left = of server] (client) {Client};
    \node[below of=server, node distance=3cm] (server_ground) {};
    \node[below of=client, node distance=3cm] (client_ground) {};
    \draw (client) -- (client_ground);
    \draw (server) -- (server_ground);
    
    \draw[->] ($(client)!0.10!(client_ground)$) -- node[sloped,above,midway]{\tiny\texttt{Initial(0): CRYPTO(CH)}} ($(server)!0.20!(server_ground)$);
    \draw[<-] ($(client)!0.40!(client_ground)$) -- node[sloped,above,midway]{\tiny\texttt{Initial(0): CRYPTO(SH), ACK(0)}} ($(server)!0.30!(server_ground)$);
    
    
    \draw[<-] ($(client)!0.45!(client_ground)$) -- node[sloped,below,midway]{\tiny\texttt{Handshake(0): CRYPTO(EE, CERT, CV, FIN)}} ($(server)!0.35!(server_ground)$);
    
    \draw[->] ($(client)!0.60!(client_ground)$) -- node[sloped,above,midway]{\tiny\texttt{Initial(1): ACK(0)}} ($(server)!0.70!(server_ground)$);
    \draw[->] ($(client)!0.65!(client_ground)$) -- node[sloped,below,midway]{\tiny\texttt{Handshake(0): CRYPTO(FIN), ACK(0)}} ($(server)!0.75!(server_ground)$);
    
    \draw[<-,dashed] ($(client)!0.95!(client_ground)$) -- node[sloped,below,midway]{\tiny\texttt{1-RTT(0): CRYPTO(NST)}} ($(server)!0.85!(server_ground)$);
    

    \end{tikzpicture}
    \caption{1-RTT connection in our \texttt{handshake} test.}
    \label{fig:ietf-quic-connection-1-rtt}
\end{figure}
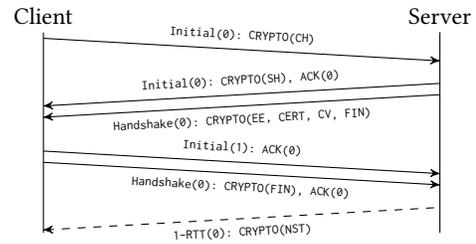

\begin{figure}
	\centering
	\includegraphics[width=\columnwidth]{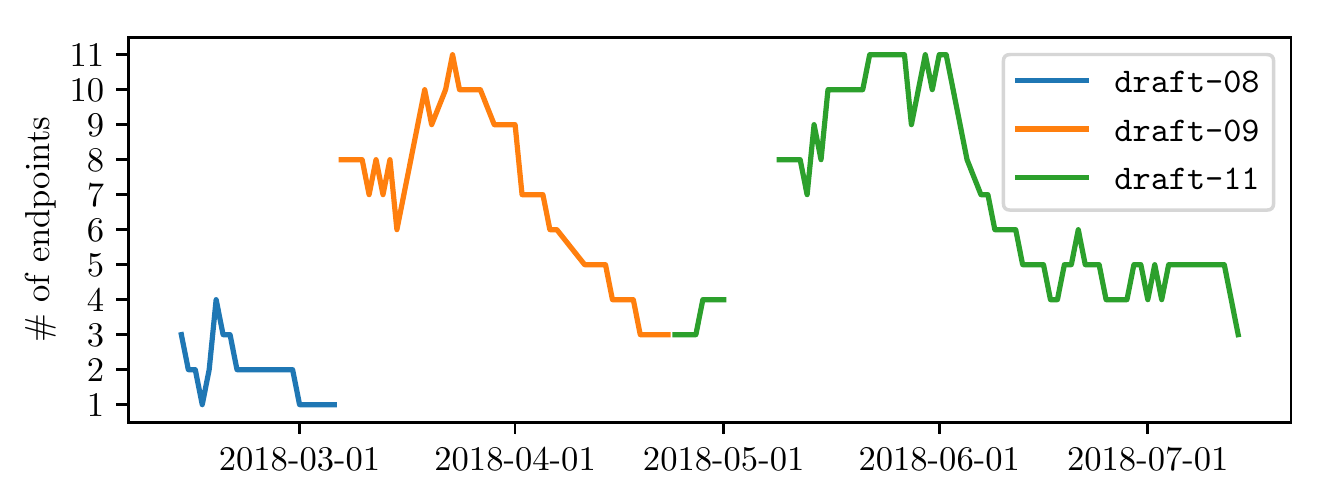}
	\caption{Number of endpoints succeeding our \texttt{handshake} test.}
	\label{fig:scenario-handshake-success}
\end{figure}

\textbf{Successful handshakes.}
Version negotiation and connection establishment being separate mechanisms, there could be a mismatch between the number of implementations that announce a particular version and the number that effectively support it. To investigate this possibility, we report the number of implementations that successfully performed a 1-RTT handshake with our test suite. It is based on the data collected by the \texttt{handshake} scenario which discerns various causes of 1-RTT handshake failure.
Figure~\ref{fig:ietf-quic-connection-1-rtt} illustrates the behaviour of our \texttt{handshake} test. We can observe that the test performs a complete 1-RTT handshake and derives the corresponding session keys. The server can send a \textit{New Session Ticket} (NST) which will be decrypted by the test.

Figure~\ref{fig:scenario-handshake-success} reports the number of endpoints that succeed our handshake test over the 6-month period.
During this period, we implemented \texttt{draft-08}, \texttt{draft-09} and \texttt{draft-11} of the specification. We chose to not deploy \texttt{draft-10}, because most implementers indicated that they were willing to support the next version as soon as possible\,\cite{quicdev-slack}. \texttt{draft-10} contained very few new features when compared to \texttt{draft-11}. We can indeed observe in Figure~\ref{fig:draft-versions} that only a maximum of four implementations simultaneously announced its support.

Overall, the resulting graph contains several slight fluctuations when compared to Figure~\ref{fig:draft-versions}. These fluctuations are a result of the rapid pace at which changes are deployed amongst all the tested implementations. Some of these changes have caused interoperability problems. This is expected as implementing the version negotiation involves simpler mechanisms than the 1-RTT handshake.

\begin{figure}
	\centering
	\includegraphics[width=\columnwidth]{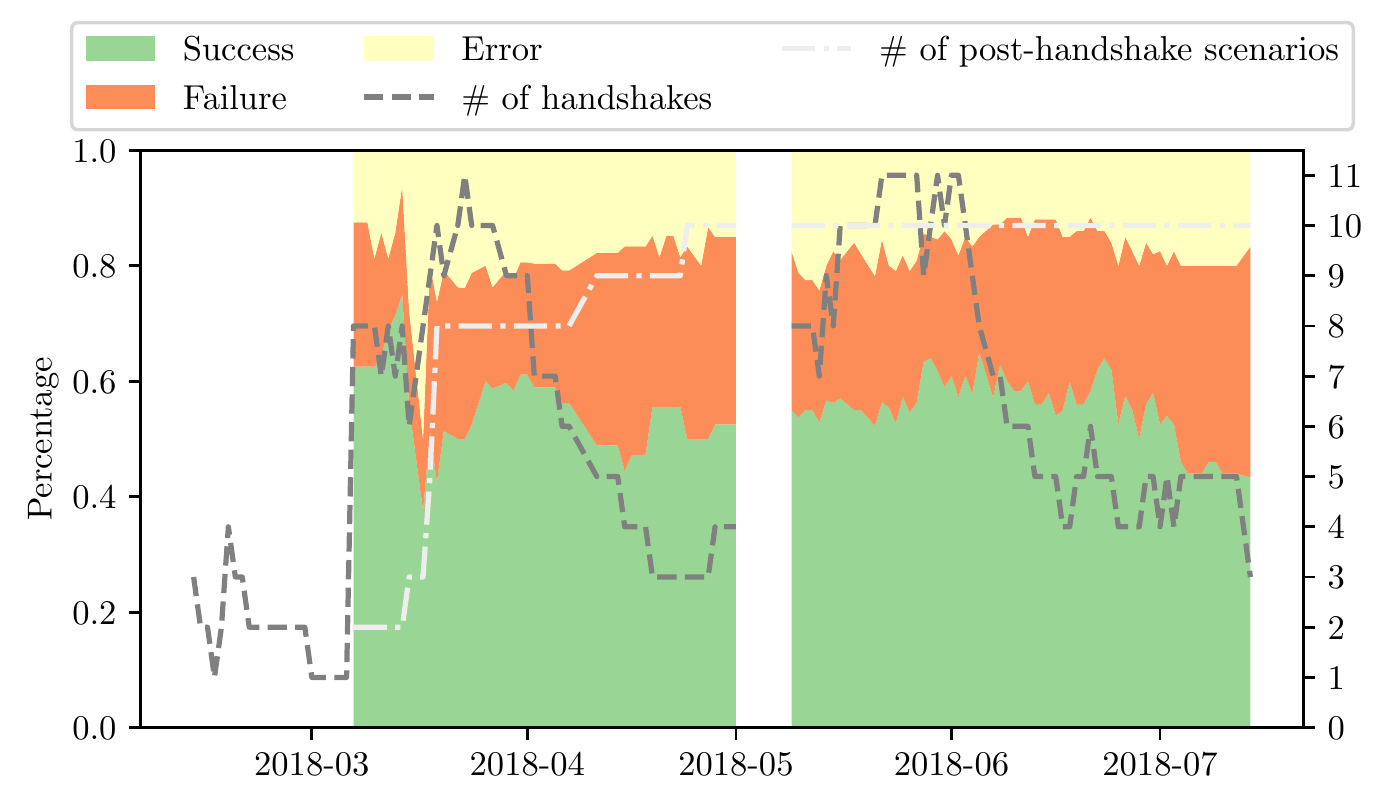}
	\caption{Percentage of outcomes for post-handshake tests.}
	\label{fig:percentage}
\end{figure}

\textbf{Test suite outcome percentage.}
We present the evolution of the success, failure and error rates over the entire test suite during our data collection period in Figure~\ref{fig:percentage}. We computed the percentage over the tests that require the handshake to complete and only kept the implementation that succeeded this handshake. We overlay the number of these implementations as well as the number of these tests on this figure. \textit{Success} corresponds to a successful execution of a test. \textit{Failure} reports the tests that were violated and \textit{Error} reports the tests for which a prerequisite was missing, e.g. the endpoint crashed, no IPv6 address could be resolved, no unidirectional streams were available, \ldots

We can note that most of the fluctuations occurred during March 2018 when most of the tests were introduced. Based on the feedback of implementers\,\cite{quicdev-slack}, we updated several tests to improve their correctness and address false positives.

The Figure illustrates that the \textit{Success}-\textit{Failure} ratio follows the number of implementations available. The most active ones rapidly move from one version to the next one and positively impact this ratio. Implementations that are slower to evolve usually have a lower ratio. Finally, the curves show many fluctuations that are indicative of the dynamic nature of these QUIC implementations.

\subsection{Case studies}

We review in this section some test scenarios which reported bugs in several implementations. For each test, we first explain its intent, then we report the evolution of its results based on the feedback submitted to and received from the implementers. We concentrate on a 3-month period starting on the 1st of March 2018.

\begin{figure}
    \centering
    \begin{tikzpicture}[node distance=0.75\columnwidth,auto,>=stealth']
    \node[] (server) {Server};
    \node[left = of server] (client) {Client};
    \node[below of=server, node distance=4cm] (server_ground) {};
    \node[below of=client, node distance=4cm] (client_ground) {};
    \draw (client) -- (client_ground);
    \draw (server) -- (server_ground);
    
    \draw[->] ($(client)!0.10!(client_ground)$) -- node[sloped,above,midway]{\tiny\texttt{   Initial(0): CRYPTO(CH, initial\_max\_stream\_data\_bidi\_local=80)}} ($(server)!0.20!(server_ground)$);
    
    \draw[dashed,color=white] ($(client)!0.11!(client_ground)$) -- node[midway,below,color=black]{$\cdots$} ($(server)!0.21!(server_ground)$);
    
    
    \draw[->] ($(client)!0.20!(client_ground)$) -- node[sloped,below,midway]{\tiny\texttt{1-RTT(0): STREAM(0, "GET /index.html\textbackslash{r}\textbackslash{n}", 17 bytes)}} ($(server)!0.30!(server_ground)$);
    
    \draw[<-] ($(client)!0.50!(client_ground)$) -- node[sloped,below,midway]{\tiny\texttt{1-RTT(0): STREAM(0, "<html>...", 80 bytes), ACK(0)}} ($(server)!0.40!(server_ground)$);
    
    \draw[->] ($(client)!0.60!(client_ground)$) -- node[sloped,below,midway]{\tiny\texttt{1-RTT(1): MAX\_STREAM\_DATA(0, 160 bytes), ACK(0)}} ($(server)!0.70!(server_ground)$);
    
    \draw[<-] ($(client)!0.90!(client_ground)$) -- node[sloped,below,midway]{\tiny\texttt{1-RTT(1): STREAM(0, "...</html>", 80 bytes), ACK(1)}} ($(server)!0.80!(server_ground)$);

    \end{tikzpicture}
    \caption{Example flow for our \texttt{flow\_control} test.}
    \label{fig:flow-control-network}
\end{figure}
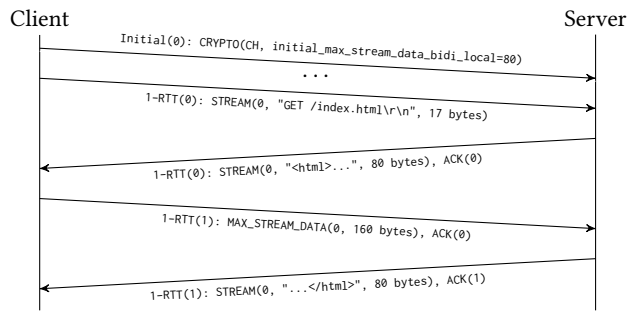
\textbf{Flow control.} Flow control is an important part of a transport protocol. It prevents a fast sender from overwhelming a slow receiver. A peer can signal flow control through two different mechanisms in QUIC. The first are the transport parameters. For instance, parameter \texttt{initial\_max\_stream\_data\_bidi\_local} allows a client to limit the amount of data that a server can send on a stream initiated by the former. The second is the \texttt{MAX\_STREAM\_DATA} frame, which advertises higher limits.

The \texttt{flow\_control} test, as illustrated in Figure~\ref{fig:flow-control-network}, initiates a connection and sets the \texttt{initial\_max\_stream\_data\_bidi\_local} parameter to 80 bytes. This limit has been chosen sufficiently low to be smaller than most of the web pages that are hosted by the endpoints. Once the connection is established, the test sends an HTTP request and waits for the server to send the first 80 bytes of the response. The server must not send more than 80 bytes because of the limit imposed by the transport parameter. Once these bytes are received, the test sends a \texttt{MAX\_STREAM\_DATA} frame that raises the limit to 160 bytes. The test ensures that the server resumes the sending of data after receiving the frame.

We found several implementations failing this test at different stages of the specification process. On the 10th and 17th of March 2018 two implementations entered a loop when running the \texttt{flow\_control} scenario. We reported these bugs and discussed with their implementers. The first was repeatedly sending \texttt{ACK} frames due to an incorrect integration of flow control with other mechanisms. The second was sending empty \texttt{STREAM} frames, which is forbidden by the specification, because of a missing corner-case when clamping these frames according to flow control. The test results collected after the 20th of March indicated that both implementers had fixed the bug.

We found another implementation that incorrectly implemented flow control on the 23rd of March. It only divided its response into two pieces, the first being 80-bytes long. We reported the bug and were notified that a fix was implemented, which was confirmed by the test results shortly after. On the 18th of May 2018, after this implementation added support for \texttt{draft-11}, we observed a regression regarding this test. The implementation aggressively sent \texttt{STREAM\_BLOCKED} frames and retransmissions of the second half of data requested. We did not observe the deployment of a fix before the end of our data collection. We later learned that its developer was not active any more during this period.

\textbf{Stream transitions reordering.} A QUIC implementation must be able to react appropriately when packet reordering occurs. We can discern two cases which can induce packet reordering. The first is introduced by the use of different network paths, due to, e.g. load balancers.
The second one is caused by a packet loss during the transmission of a series of packets. The data of the lost packet will be retransmitted and received after the rest of the series.

The \texttt{stream\_opening\_reordering} test simulates the first type of reordering. It initiates a connection and then sends an HTTP request in two packets. The first packet contains the graceful closure of the client-side of the request stream. The second contains the data of the stream, which contains the request. The first packet is sent with a higher packet number than the second packet. The test successfully completes once the server has responded to the request.

We report three of the cases we observed during the 3-month period. The first one lead to a one-to-one conversation with an implementer. The scenario triggered a livelock in their implementation and the latter did not produce any kind of observable external behaviour. We provided assistance to install the test suite and run it against a local and better-instrumented version of the implementation.

On the 11th of May, we detected a regression for a particular implementation for which support of \texttt{draft-11} was recently added. We were not actively analysing the data on this day and thus did not report the bug. We later found that the implementer had consulted the test result and fixed the bug. We argue that this is an indication of the benefits of an autonomous test suite that runs on a daily basis and provides public results.

Finally, this test triggered a bug in the \texttt{ACK} frame generation of an implementation on the 22nd of May. We believe that the bug was discovered shortly after its introduction, as the results from the past days did not reveal it. The bug caused the generated \texttt{ACK} frame to report $2^{64} - 1$ missing packets, probably due to an overflow induced by the reordering of packets. Indeed, considering that one could determine the gap between the received packet and the last received packet by subtracting the received packet number with the last received packet number, reordering then causes an overflow. The implementation source code not being public, we could not confirm this hypothesis.

\subsection{Results grid}

\begin{figure}
	\centering
	\includegraphics[width=\columnwidth]{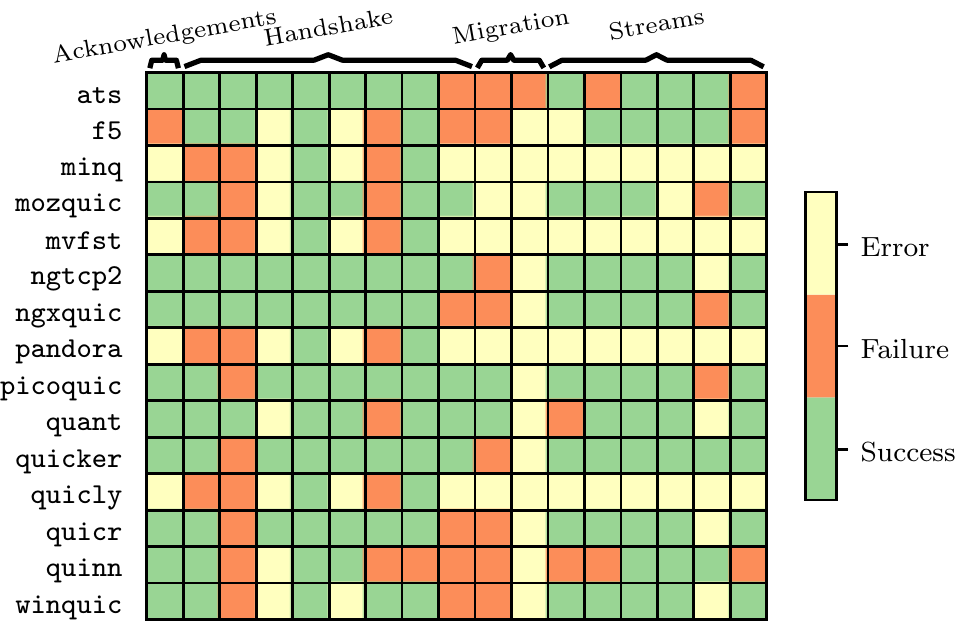}
	\caption{Results grid on the 1st of June 2018.}
	\label{fig:draft-11-grid}
\end{figure}

We conclude this section by presenting Figure~\ref{fig:draft-11-grid} which summarises the outcomes for the different tests and QUIC implementations. The grid is a snapshot captured on the 1st of June 2018.

Those results show the diversity of the outcomes generated by the available implementations.
We can observe that three of them, i.e.\ \texttt{minq}, \texttt{pandora} and \texttt{quicly} only succeeded two scenarios. These scenarios only collect metrics and do not enforce requirements when the IUT is unavailable. We can also note that most of the runs of the two connection migration tests were either unsuccessful or could not execute. While the mechanisms tested were part of \texttt{draft-11}, the implementers did not include them in the corresponding \textit{Implementation Draft}\,\cite{5th-implementation-draft}.

\section{Discussion}\label{sec:discussion}

In this paper we have proposed a first active test suite for the QUIC protocol based on the current IETF specification. We detailed its architecture and the supported test scenarios. We presented the results collected using this test suite and reported two case studies. The test suite has already been used by the QUIC community.
Its source code is publicly available under an open-source licence\footnote{See \url{https://github.com/QUIC-Tracker}}. Two implementers have already integrated the test suite as part of their workflow, independently of our public instance.

The tool being open also implies that it can be reused, extended and improved by the QUIC community. Due to the very modular design of our architecture, it can be extended in different ways. New scenarios can be implemented to cover new features of the protocol and collect new metrics. The QUIC toolbox can be reused for other purposes. It can also be extended with new features that are not currently supported, such as sending coalesced packets, or improved with a better user-facing API.
The visualisation application can also be improved, e.g. by adding more feedback to the implementers based on the trace format. We note that one is not required to use this web application, and can instead consume the test results in the traces using other applications. For instance, we developed a set of scripts that allows generating CSVs based on these traces, which were used to produce Figure~\ref{fig:draft-versions} and Figure~\ref{fig:scenario-handshake-success}\footnote{\url{https://github.com/QUIC-Tracker/web-app/tree/master/quic_tracker/postprocess}}.

We plan to continue to update the test suite to track the evolution of the IETF specification and later to detect how QUIC server implementations have been tuned with heuristics for, e.g.\ retransmissions and congestion control schemes.
However, we limit our approach only to QUIC servers.
The protocol also requires the compliance of QUIC clients to the specification.
Including them in our study would raise several challenges beyond the additional implementation efforts.

\textbf{How to initiate connections from clients to the test tool ?}
We chose a black-box approach as the source code of QUIC implementations is not always available. Applying this approach to client testing requires some techniques to encourage diverse clients to connect to the test tool.

\textbf{How to identify the various implementations connecting to the test tool ?}
In the server-side approach, we know to which servers the test tool connects to. If many clients can anonymously connect to the test tool, correctly identifying the tested implementations becomes critical for providing relevant feedback to the QUIC implementers.

\textbf{Which QUIC clients are widely-deployed today ?}
While several server implementations are known thanks to their participation in the IETF interoperability tests, there is no equivalent for clients as of this writing. However, this is likely to change as the QUIC specification is finalised.

We intend to include QUIC clients to our approach in the future and we hope to be able to capture the full diversity of the emerging QUIC ecosystem, with as many interesting behaviours as we presented in this study.

\section*{Acknowledgements}

We would like to thank the QUIC implementers and the participants to the IETF Hackathon in London who provided feedback on the test suite. This work was partially supported by funding from the Walloon Government (DGO6) within the MQUIC project.

\newpage

\bibliographystyle{ACM-Reference-Format}
\bibliography{reference}

\newpage

\end{document}